# TRANSCRIPTION FACTOR BINDING SITE PREDICTION WITH MULTIVARIATE GENE EXPRESSION DATA

By Nancy R. Zhang, Mary C. Wildermuth[1] and Terence P. Speed[1]

*Stanford University, University of California, Berkeley and University of California, Berkeley*

Multi-sample microarray experiments have become a standard experimental method for studying biological systems. A frequent goal in such studies is to unravel the regulatory relationships between genes. During the last few years, regression models have been proposed for the *de novo* discovery of *cis*-acting regulatory sequences using gene expression data. However, when applied to multi-sample experiments, existing regression based methods model each individual sample separately. To better capture the dynamic relationships in multi-sample microarray experiments, we propose a flexible method for the joint modeling of promoter sequence and multivariate expression data.

In higher order eukaryotic genomes expression regulation usually involves combinatorial interaction between several transcription factors. Experiments have shown that spacing between transcription factor binding sites can significantly affect their strength in activating gene expression. We propose an adaptive model building procedure to capture such spacing dependent *cis*-acting regulatory modules.

We apply our methods to the analysis of microarray time-course experiments in yeast and in Arabidopsis. These experiments exhibit very different dynamic temporal relationships. For both data sets, we have found all of the well-known *cis*-acting regulatory elements in the related context, as well as being able to predict novel elements.

**1. Problem formulation and review of methods.** Two important sources of high-throughput data have become available to modern biology: microarrays, which allow measurement of genome-wide gene expression patterns over multiple biological conditions, and sequenced genomes, which allow computational search of any sequence pattern of interest in any genomic region of

---

Received April 2007; revised September 2007.

[1]Supported by NSF Arabidopsis 2010 Grant MCB-04-0267.

*Key words and phrases.* Multivariate analysis, linear models, transcription regulation, DNA motifs, gene expression.

interest. This wealth of resources has triggered attempts to computationally learn the regulatory grammar within the promoters of genes through combined analysis of expression and promoter sequence data. The logic that underlies such efforts is that gene expression is initiated by the binding of transcription factors to upstream sequences (called *cis*-acting regulatory elements), and thus, gene expression patterns should be correlated with the presence of certain *cis*-acting regulatory elements in the promoter sequence.

Not long after the advent of microarray technology, gene expression data was used to locate known or putative transcription factor binding sites (TFBS). Initially, most methods were based on the prespecification of a set of hypothesized co-expressed genes, which can be obtained through cluster analysis of gene expression profiles. Statistical methods have been developed to find enriched motifs in the promoters of such pre-specified gene sets [Bailey and Elkan (1994) and Liu, Brutlag and Liu (2002)]. Such methods can model degenerate motifs through position specific weight matrices or graph-based motif representations [Fratkin et al. (2006)]. Models that incorporate interactions between motifs have also been developed [Zhou and Wong (2004)]. However, all of these methods are very sensitive to the prespecified gene list. It is usually unclear how to obtain such a gene list from microarray data, because clustering methods are often unreliable and most genes are not part of a tight cluster. Also, these methods make use of the expression data only to the extent of obtaining the set of hypothesized coregulated genes, and do not explicitly model the relationship between promoter sequence content and gene expression.

Regression based approaches have been proposed to more directly model the relationship between the expression pattern of genes and the repertoire of motifs in their promoters. Bussemaker, Li and Siggia (2001) proposed the following simple linear model between $X_{g,m}$, the count of a motif $m$ in the promoter of gene $g$, and $Y_g$, the expression of gene $g$:

$$Y_g = \sum_m X_{g,m}\beta_m + \epsilon_g,$$

where the summation is over the set of all motifs that is believed to contribute to the expression of genes in the sample. This basic model has since been expanded to utilize position specific weight matrices (PSWM) instead of counts [Conlon et al. (2003) and Das, Nahlé and Zhang (2006)], as well as to model interactions between motifs that appear together in the same promoter sequence [Das, Nahlé and Zhang (2004) and Keles, Van der Laan and Vulpe (2004)]. Notably, Keles, Van der Laan and Vulpe (2004) used logic regression to model combinatorial motif interactions and Das, Nahlé and Zhang (2004) used linear splines to capture the hypothesized log-sigmoidal relationship between transcription response and motif strength. All of the methods so far mentioned have attained some degree of success with data from yeast, but with the exception of Das, Nahlé and Zhang (2006), the applicability of these methods in higher order organisms is still unproven.

At present, all of the existing regression based methods model only univariate expression data. For example, Bussemaker, Li and Siggia (2001), Das, Nahlé and Zhang (2004), Keles, Van der Laan and Vulpe (2004) and Conlon et al. (2003) all analyzed the yeast cell cycle time series [Spellman et al. (1998)] by doing a separate regression at each time point. However, it is also clear from these studies that, for the cell cycle experiments, the known biological motifs have a meaningful time-varying pattern, while the false positive motifs have a pattern across time that is not recognizable or that is typical of experimental artifacts. Multivariate gene expression data, measured across different biological conditions, times and treatments, convey much more information than single chip data. However, at present there is no clear method to combine information across samples. It is often unclear, when multiple samples are available, how to quantify whether a gene's expression profile is "interesting." Yet, to capture dynamic regulatory relationships and to respond to the growing availability of multi-chip data, it is necessary to combine information across samples in modeling *cis*-acting regulatory elements. In this paper we present a linear model for this purpose.

We also propose in this paper a new framework for modeling interactions between *cis*-acting regulatory elements that takes into consideration the distance between the elements in the promoter sequence. Transcription factors, when bound to the promoter sequence, interact with other transcription factors and nuclear proteins to initiate or inhibit transcription. The distance between the binding sites affects the strength of such interactions. For example, Rushton et al. (2002) showed using synthetic plant promoters that the distance between the binding sites of the WRKY transcription factors causes a pronounced difference in the strength of the module. Another example is given in Segal and Berk (1991), where they showed that in vivo transcription stimulation by Sp1 transcription factor binding sites in the adenovirus type 2 early region 1B promoter is strongly dependent on its distance from the TATA box. In an analysis of 4 yeast species, Chiang et al. (2003) found that pairs of jointly conserved motifs exhibit nonrandom relative spacing. In this paper we give a computational method to incorporate this effect into motif detection.

A critical step in all regression based methods is the selection of the best subset of motifs to be included in the model. The initial set of motifs under consideration can be quite large, especially for *de novo* motif finding. Most previous methods [Bussemaker, Li and Siggia (2001), Conlon et al. (2003), Keles, Van der Laan and Vulpe (2004)] resorted to simple stepwise variable addition to search the space. Das, Banerjee and Zhang (2004), Das, Nahlé and Zhang (2006), which uses the multivariate adaptive regression splines (MARS) algorithm [Friedman (1991)], used a forward stepwise search followed by model pruning. In this paper we examine two different model selection methods with the aim of obtaining a simple, interpretable model. We use a permutation study to evaluate the effectiveness of our model selection procedure and to estimate the expected number of false positives. Most previous studies on this problem have not done such specificity evaluations, which we believe to be critical to the understanding and assessment of models.

The paper is organized as follows: To motivate our methods, we begin by describing two different multi-sample gene expression experiments in Section 2. We present our model and methods in Section 3. Methods of statistical validation are important for interpretation of the results, and thus, in Section 4 we describe three validation methods, including a new approach based on flanking sequence information content that has never been applied in previous published studies. In Section 5 we present the results for the experiments described in Section 2. Finally, we conclude with a few remarks in Section 6.

**2. Description of experiments.** We begin by describing two multi-sample gene expression experiments which will be used to illustrate our methods. Both are time-course experiments. The first is a periodic time course representing the self-regenerating process of cellular growth and division. The second is a non-periodic time course representing an organism's response to a biotic stimulus (a pathogen).

2.1. *Yeast cell cycle.* The cell cycle is a tightly coordinated set of processes by which cells grow, replicate their DNA, segregate their chromosomes, and divide into daughter cells. Checkpoints during the cycle ensure that at specific time points, specific processes must have been completed. Such coordination requires a complex network of regulatory relationships. We apply our methods to learn these relationships from a set of gene expression measurements captured at time-points during the cell cycle. The data set we use comes from the $\alpha$-factor synchronized cultures from Spellman et al. (1998), and can be downloaded from the website **http://cellcycle-www.stanford.edu**.

The samples of the $\alpha$-arrest experiment are taken at 7 minute intervals after synchronization of the cell culture. There are a total of 18 samples, spanning two cell cycles. Thus, many genes related to the cell cycle have a periodic expression with two periods captured in this data set. Much is known about the regulatory mechanisms involved, and a list of the transcription factors that is known to play a crucial role can be found in Spellman et al. (1998). For the yeast analyses, we included 1600 genes: 800 defined by Spellman as cell-cycle related genes [Spellman et al. (1998)] and 800 genes selected at random from genes that did not exhibit a cell-cycle related pattern. In this way we are able to determine whether *cis*-acting regulatory elements associated with specific gene expression patterns emerge from a mixed dataset. Details about data pre-processing are given in Section A.1.

2.2. *Systemic acquired response in plants.* Plant are commonly exposed to bacterial, fungal and viral pathogens. In response to certain pathogens, plants activate defensive responses resulting in enhanced resistance to subsequent pathogen exposure. This acquired resistance response can be both local and systemic (occurring in uninfected parts of the plant). The small molecule 2-hydroxybenzoic, commonly known as salicylic acid (SA), is required for this local and systemic resistance response. Wildermuth et al. (2001) showed that this SA is synthesized via isochorismate synthase (At-ICS1) in Arabidopsis thaliana. Null *ics*1 mutant plants do not synthesize SA in response to pathogen, are more susceptible to pathogens and compromised in local and systemic acquired resistance (SAR) responses.

To investigate the specific components and processes of plant defense mediated by SA, Wildermuth et al. (2007) infected wild type and *ics*1 mutant Arabidopsis plants with the powdery mildew fungal pathogen Golovinomyces orontii. In this replicated time-course experiment, samples were harvested at 0 (just prior to infection), 6, 24, 48, 72, 120 and 168 hrs post infection (hpi). This timing focuses on the progressive growth and reproduction of the fungus. Analysis of this ATH1 Affymetrix dataset indicates that the majority of genes with a significant difference in expression in the *ics*1 mutant compared with wild type exhibit an increase in expression in response to the powdery mildew in wild type plants with abolished or reduced expression in the mutant. A number of these are known to act downstream of SA in SAR. In addition, there are genes that exhibit enhanced expression in the *ics*1 mutant vs. wild type in response to powdery mildew. In our analysis of this data, we would like to capture this dynamic time-relationship in finding regulatory TFBS related to SA-mediated SAR. The details of the regulatory networks mediating SAR are an area of active investigation. Binding sites for a few key transcription factors involved in plant defense have been determined; however, much is still unknown. For the Arabidopsis analysis, we included the top 1500 genes that exhibited differential expression between the wild type and *ics*1 mutant in response to pathogen and the bottom 1500 genes exhibiting no difference in expression between wild type and mutant. The inclusion of an equal subset of genes with unaltered expression serves as a control, allowing us to determine if we are able to identify genes specifically associated with altered expression patterns. Details of the data pre-processing are given in Section A.1.

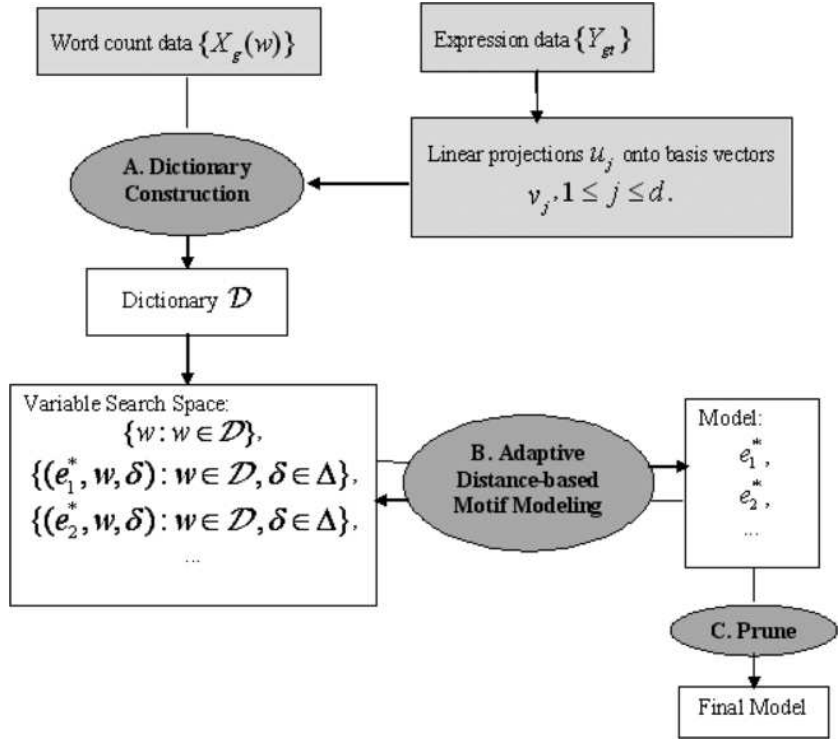

Fɪɢ. 1.   *Flow diagram of the analysis process. We start with the multivariate expression data for a filtered set of genes and the counts of all nondegenerate motifs of pre-specified lengths in the upstream promoter sequence of these genes. Linear contrasts* $\mathbf{u}_j$ *of the expression data are used to build the dictionary (step* A*), and subsequently build the model (step* B*). The model is then pruned (step* C*) for parsimony. The steps labeled* A, B *and* C *are detailed in Sections 3.2, 3.3 and 3.4, respectively.*

**3. Model and method.**   Figure 1 shows a flow-diagram of our proposed method. The main steps of the analysis follow: (A) Dictionary construction, (B) Adaptive distance-based model building, and (C) Model pruning. We start in Section 3.1 with a description of our model and loss function, which will be integral to steps (B–C).

3.1. *Model and loss function specification.*   The observed data are $\mathbf{Y} \overset{\text{def}}{=} \{Y_{g,t} : 1 \le g \le G, 1 \le t \le T\}$, where $Y_{g,t}$ is the $\log_2$ expression intensity for gene $g$ in sample $t$, and $\mathbf{S} \overset{\text{def}}{=} \{S_{g,i} : 1 \le g \le G, 1 \le i \le R\}$, where $S_{g,i}$ is the DNA base at position $-i$ from the start of gene $g$ in the $5'$ to $3'$ template DNA strand, which we refer to as gene $g$'s promoter. More specific definitions of the promoter sequence of a gene are given in Appendix A.1. We denote the expression and promoter sequence of a gene $g$ respectively by $\mathbf{Y}_g = \{Y_{g,1}, \ldots, Y_{g,T}\}$ and $\mathbf{S}_g = (S_{g,1}, \ldots, S_{g,R})$.

Our model assumes that $\mathbf{Y}_g$ is a sum of two orthogonal components:

$$\mathbf{Y}_g = \mathbf{Z}_g + \boldsymbol{\varepsilon}_g, \tag{1}$$

where $\mathbf{Z}_g$ is the signal that the experiment is designed to measure, and $\boldsymbol{\varepsilon}_g$ is the error component due to technical or biological noise. Note that $\boldsymbol{\varepsilon}_g$ may be systematic error with a biologically meaningful trend in $t$, but that is not of interest in the context of the given experiment. By orthogonality of $\boldsymbol{\varepsilon}_g$ and $\mathbf{Z}_g$, we mean that the covariance matrix $\mathrm{Cov}[\mathbf{Z}_g, \boldsymbol{\varepsilon}_g] = \mathbf{0}$.

We model the dependence of $\mathbf{Y}_g$ on $\mathbf{S}_g$ through the signal component $\mathbf{Z}_g$. Specifically, let $\mathbf{Z}_g$ reside in a $d$ dimensional linear subspace of $R^T$, with linear decomposition

$$\mathbf{Z}_g = \sum_{j=1}^{d} u_{j,g} \mathbf{v}_j \tag{2}$$

on basis vectors $\mathbf{v}_1, \ldots, \mathbf{v}_d$. We assume the following linear model:

$$u_{j,g} = \beta_{0,j} + \sum_{e \in \mathcal{E}} \beta_j(e) X_g(e) + \epsilon_{j,g}, \qquad j = 1, \ldots, d, \tag{3}$$

where $\mathcal{E}$ is a set of promoter elements. Section 3.3 defines the concept of promoter element in detail, while here we simply describe it intuitively as a collection of DNA letters that satisfy a certain arrangement. The variables $X_g(e)$, defined in Section 3.3, quantify the presence of promoter element $e$ in $\mathbf{S}_g$. We assume that the errors $\epsilon_{j,g}$ are independent across $g$ and $j$ with mean 0 and equal variance.

Note that in the model defined in (3), the set of promoter elements $\mathcal{E}$ is common to all basis functions $\{\mathbf{v}_j : j = 1, \ldots, d\}$. Alternatively, one could fit a separate model with a different set $\mathcal{E}$ for each basis function. This would be meaningful if the basis functions were contrasts that individually represent a quantity of interest. However, with model (3), we are looking for a set of promoter elements that play a significant role in the systemic context of the experiment (represented by the signal component $\mathbf{Z}_g$), without being specific to the choice of basis. The goal is then to choose $\mathcal{E}$ and parameters $\boldsymbol{\beta}_{\mathcal{E}} = \{\beta_j(e) : e \in \mathcal{E}, 1 \le j \le d\}$ that are the most effective in explaining the variance in $\mathbf{Z}_g$. For each gene $g$, assume that we have the fitted values $\{\hat{u}_{j,g} : j = 1, \ldots, d\}$, which gives us $\hat{\mathbf{Z}}_g = \sum_{j=1}^{d} \hat{u}_{j,g} \mathbf{v}_j$, the fitted value for $\mathbf{Z}_g$. The loss function is defined as

$$L(\mathcal{E}, \boldsymbol{\beta}_{\mathcal{E}}) = \sum_{g=1}^{G} \|\hat{\mathbf{Z}}_g - \mathbf{Z}_g\|^2. \tag{4}$$

If $\mathcal{E}$, the set of promoter elements in the model, were fixed, the goal would be to find $\boldsymbol{\beta}_{\mathcal{E}}$ that maximizes (4). However, for varying $\mathcal{E}$, the loss function would of course need to be penalized for the degrees of freedom of the model specified by $\mathcal{E}$. Section 3.4 discusses possible penalty functions.

In practice, one would need to specify the basis functions $\{\mathbf{v}_j\}$ of the signal component $\mathbf{Z}_g$. This can sometimes be done a priori by using contextual knowledge of the experiment. For example, for case-control experiments, $\mathbf{v}_j$ can be set to the contrasts of interest. However, for the two experiments that we study, there are no obvious contrasts. For the yeast cell cycle experiment, an a priori meaningful set of basis could be orthogonal periodic functions with peaks at different phases of the cycle. For the powdery mildew infection experiment, a good choice of basis is even harder to find without examination of the data.

A good method for choosing $\{\mathbf{v}_j\}$ is to use the principal components of $\mathbf{Y}$, which proved effective on both the yeast and the Arabidopsis experiments. With some overlap with previous notation, consider the singular value decomposition $\mathbf{Y} = \mathbf{U}\mathbf{\Lambda}\mathbf{V}'$, where we let $\mathbf{U}$ be the matrix of score vectors $[u_1, \ldots, u_T]$, $\Lambda$ be the diagonal matrix with diagonal elements $[\lambda_1, \lambda_2, \ldots, \lambda_T]$, and $V = [v_1, \ldots, v_T]$ be the $T \times T$ square matrix that contains the loadings of the $T$ principal component vectors. Often, a visual examination of the loadings can yield meaningful linear projections of the data, and good choices for $\{v_j\}$. Using contextual knowledge about the experiment, one may be able to identify a subset $\mathcal{A} \subseteq \{1, 2, \ldots, T\}$ of principal components vectors to serve as the basis. Otherwise, one can choose the top few principal components using a scree plot. With the basis vectors chosen using principal components, the loss function (4) would then be equivalent to the weighted sum of losses over the principal component scores:

$$(5) \qquad L_{\mathcal{A}}(\mathcal{E}, \boldsymbol{\beta}_{\mathcal{E}}) = \sum_{j \in \mathcal{A}} \lambda_j^2 \|u_j - \boldsymbol{\beta}_j X_{\mathcal{E}}\|^2,$$

where $\boldsymbol{\beta}_j = (\beta_j(e) : e \in \mathcal{E})$. Note that the weight of each component in the above loss function is exactly the variance of the data along that component.

If $\mathcal{A}$ were chosen to be the entire set of all $T$ principal components, then the minimum of $\mathcal{L}_{\mathcal{A}}(\mathcal{E}, \boldsymbol{\beta}_{\mathcal{E}})$ over $\boldsymbol{\beta}_{\mathcal{E}}$ would be equivalent to the naive unweighted sum of the squared error losses from a separate regression for each sample:

$$\min_{\boldsymbol{\beta}_{\mathcal{E}}} \mathcal{L}_{\mathcal{A}}(\mathcal{E}, \boldsymbol{\beta}_{\mathcal{E}}) = \sum_{j=1}^{T} \min_{\beta_j(\mathcal{E})} \|Y_{.,j} - \beta_j(\mathcal{E})X_{.,\mathcal{E}}\|^2.$$

In model (1), this would be equivalent to assuming $\boldsymbol{\varepsilon}_g = 0$ for all genes $g$. By selecting $\mathcal{A}$ to be a proper subset of $\{1, \ldots, T\}$, we are performing weighted multivariate response regression on a reduced dimensional subspace of $Y$.

For time-course experiments, this method finds motifs that most significantly affect the shape of the time course. In high dimensions many shapes are possible. We focus on the shapes that are most meaningful to the experiment, or that are the most effective in explaining the data, or both. The model gains power over previous methods by combining information across time-points to capture the dynamic relationships that can only be observed across samples. In particular, this multivariate cross-sample approach is preferable to the single-sample approach for finding motifs related to pathways that have distinct time-related patterns of activity.

The principal components approach gives an automatic, data-based method of weighting the residual sum-of-squares when multiple basis vectors (components) are chosen. In both the yeast cell cycle and the Arabidopsis experiments, we found more than one meaningful orthogonal projection of the data. One could, in principle, apply the methods described in the next few sections to each projection separately, obtaining separate sets of motifs $\mathcal{E}$. However, with the weighted loss function (4), we hope to capture motifs that may not be strongly correlated with any single projection, but that are weakly correlated with many projections. Another benefit of combined analysis of multiple orthogonal projections is in the detection of interactions between TFBS. Transcription factors that are active in different pathways may interact to assume a new role. Such interactions may be lost if one limits the analysis to one-dimensional projections.

When there are multiple basis vectors, what determines which motifs get added to the model? Some insights can be gained from examining the simple case where $\mathcal{E}$ is a singleton $\{e\}$. Then, the weighted loss (5) has a simple representation

$$L_{\mathcal{A}}(e, \boldsymbol{\beta}_e) = \sum_{j \in \mathcal{A}} \lambda_j^2 (1 - \rho_{j,e}^2),$$

where $\rho_{j,e} = \text{corr}[u_j, X(e)]$. Thus, the promoter element $e$ that has the maximum weighted sum of squared correlation with the components in $\mathcal{A}$ would be added to the model first, with the weights being the proportion of variance explained by that component.

In the following sections we will assume that the basic vectors $\mathbf{v} = \{v_1, \ldots, v_d\}$ in (2) have been chosen, either through principal components or other methods. We will use the notation

$$\lambda_j = \|Y \mathbf{v}_j\|^2, \qquad \mathbf{u}_j = Y \mathbf{v}_j / \sqrt{\lambda_j}.$$

3.2. *Step* A. *Dictionary construction.* We chose to represent motifs using nondegenerate words. In the organisms that we study, a word and its reverse complement represent the same biological motif. This is because DNA is double stranded, and the appearance of the reverse complement of the word in the $5'$ to $3'$ template strand is equivalent to the appearance of the word in $5'$ to $3'$ orientation in the coding strand. For example, in the diagram below, the top strand is the template strand, the bottom is the coding strand, and `TTGAC` and `GTCAA` represent the same biological motif **presented 5′ to 3′**:

```
5'...TTGAC...3'
3'...AACTG...5'.
```

We start our analysis with the set of all unique deterministic biological motifs of a pre-chosen length $L$. The size of this set is $[4^L - 4^{L/2}]/2 + 4^{L/2}$ if $L$ is even, or $4^L/2$, if $L$ is odd. The above enumeration counts each word and its reverse complement only once.

In the dictionary construction step, this initial set of words is reduced to a much smaller set for subsequent model building. Although dictionary construction has been viewed in such studies as a crude pre-filtering step to reduce the size of the model search space, it is very important to the ensuing analysis. For a motif to be selected in the final model, it must first be included in the dictionary. Therefore, the dictionary must provide a rich enough starting set of motifs, while at the same time reducing the set of initial exhaustive list of words to a more computationally manageable set.

Here we chose to represent motifs as nondegenerate words. We also tried representing motifs as "consensus" sequences by including nondegenerate core letters with degenerate outermost letters. Using a 4 letter core, we allowed the outermost 2 (or 4) letters to vary. This "consensus" sequence approach reduced performance as it drastically increased the initial dictionary aggravating the multiple testing problem and increased dependency between the motifs. One could also use our approach to identify known TFBS PSWMs by setting the dictionary to a known collection of PSWMs, as in Conlon et al. ([2003](#)) and Das, Nahlé and Zhang ([2006](#)). However, the availability of PSWMs is limited for many organisms, including Arabidopsis. This is why we specifically developed a method that allows one to identify known and novel motifs without prior knowledge.

Let $D^{(0)}$ be the set of all words of length within a pre-chosen range. For each $w \in D^{(0)}$, let $X_g(w)$ be the count of the number of occurrences of $w$ in $\mathbf{S}_g$. Let $X(w) = [X_1(w), \ldots, X_G(w)]'$. For any set $\Gamma$ of motifs, we will denote by $X(\Gamma) = [\mathbf{1} \;\; [X(w)]_{w \in \Gamma}]$, with the vector of ones always included in the model matrix as the intercept term. We start by constructing a smaller dictionary $D^{(j)}$ for each chosen basis vector $\mathbf{v}_j$:

1. Let $m = M$, where $M$ is a pre-chosen value. Let $D^{(j)} = \varnothing$.
2. Repeat until $m = 0$:

   (a) Compute

   $$\xi(w) \leftarrow X(w) - X(D^{(j)})[X(D^{(j)})'X(D^{(j)})]^{-1}X(D^{(j)})'X(w),$$
   $$w \in D_L^{(0)},$$
   $$r \leftarrow \mathbf{u}_j - X(D^{(j)})[X(D^{(j)})'X(D^{(j)})]^{-1}X(D^{(j)})'\mathbf{u}_j.$$

(b) For each $w \in D_L^{(0)}$, let $p_w$ be the t-test p-value for the univariate regression of $r$ on $\xi(w)$. Add the $m$ words with smallest $p_w$ to $D^{(j)}$.

(c) Let $m = \lfloor m/2 \rfloor$.

This greedy stepwise filtering approach adds to the dictionary not only those words that are highly correlated with $\mathbf{u}_j$, but also those words that are highly correlated with $\mathbf{u}_j$ after accounting for the affects of the previously added words. Transcription factor binding sites are usually degenerate, and thus, the words with the smallest $p_w$ are often variations of the same TFBS. Therefore, at each step in the above algorithm, the set of words that are added to $D^{(j)}$ is usually swamped by overlapping words representing a single TFBS, thus reducing its richness. Since there is a high correlation among these overlapping words, the stepwise filtering approach mitigates this problem.

For the final dictionary, we set $\mathcal{D} = \bigcup_{j=1}^d D^{(j)}$.

3.3. *Step* B. *Adaptive distance-based motif modeling.* Let $\mathcal{D}$ be the dictionary of motifs constructed using the method described in Section 3.2. Let $\Delta = \{\delta_i\}_{i=1}^r$, where $\delta_i \in \mathcal{Z}^+$, be a set of possible ranges of interactions. We define a promoter element to be either a word from $\mathcal{D}$, or an interaction of the form $(e_1, e_2, \delta)$, where $e_1$, $e_2$ are themselves promoter elements, and $\delta \in \Delta$. We call elements that are words from $\mathcal{D}$ *simple*, and elements that contain interactions *composite*. Let $e$ be a promoter element. If $e$ is simple and of length $l$, then for any gene $g$, we define the locations of $e$ in $S_g$ as

$$A_g(e) = \{i : S_{g,i:i+l-1} = e\}.$$

If $e$ is composite, then its locations are defined recursively by

$$A_g(e) = \left\{ \frac{i+j}{2} : i \in A_g(e_1), j \in A_g(e_2), |j - i| \le \delta \right\},$$

where $e_1$ and $e_2$ can be either simple or composite. We denote by $I(\cdot)$ the indicator function for the event in its argument. Then, we define the variables $X_g(e)$, which are used as covariates in the model (3), as follows:

$$(6) \qquad X_g(e) = \begin{cases} |A_g(e)|, & e \text{ simple}; \\ I(|A_g(e)| \ge 1), & e \text{ composite}. \end{cases}$$

In words, if $e$ were a simple element, then $X_g(e)$ would simply be its count in the promoter of $g$, as done in Bussemaker, Li and Siggia (2001). If $e$ were composite, then $X_g(e)$ would be an indicator of whether it exists in the promoter of $g$. We define the order of a promoter element to be the number of interactions it contains:

$$\text{order}(e) = \begin{cases} 0, & e \text{ simple}; \\ \text{order}(e_1) + \text{order}(e_2) + 1, & e \text{ composite}. \end{cases}$$

We denote $X(e) = [X_1(e), \dots, X_G(e)]'$.

To learn the model defined in Section 3.1, we need to build the set $\mathcal{E}$ and estimate the parameters $\boldsymbol{\beta}$. The method we propose is a stepwise procedure, where at each step we search over a variable pool $V$ for the minimizer of the loss function (4), and add it to the model. $V$ is initialized to contain all elements in the dictionary $\mathcal{D}$. The model is initialized to contain only the intercept term $\{\beta_{0,j} : j = 1, \ldots, d\}$. With each addition of an element form the variable pool to the model, its interactions with all elements in the dictionary, at all distances in $\Delta$, are added to the variable pool. Thus, the variable pool adaptively expands with the model. The algorithm is described in detail below:

1. Initialize $V = D$, $\tilde{\mathcal{E}} = \varnothing$,
2. Repeat until $|\tilde{\mathcal{E}}| = M$:

    (a) Compute

$$r_j \leftarrow r_j - X(\tilde{\mathcal{E}})[X(\tilde{\mathcal{E}})'X(\tilde{\mathcal{E}})]^{-1}X(\tilde{\mathcal{E}})'r_j, \qquad j = 1, \ldots, d;$$

$$\xi(e) \leftarrow X(e) - X(\tilde{\mathcal{E}})[X(\tilde{\mathcal{E}})'X(\tilde{\mathcal{E}})]^{-1}X(\tilde{\mathcal{E}})'X(e), \qquad e \in V,$$

    where $X(\tilde{\mathcal{E}})$ is a matrix containing columns $\{X(e) : e \in \tilde{\mathcal{E}}\}$.

    (b) Select $e^* \in V$ by the criterion

$$e^* = \arg\min_{e \in V} \sum_{j=1}^{d} \lambda_j^2 \min_{\beta} \|r_j - \beta\xi(e)\|^2.$$

    (c) $\tilde{\mathcal{E}} \leftarrow \tilde{\mathcal{E}} \cup \{e^*\}$.
    (d) if $\operatorname{order}(e^*) < o_{\max}$, $V \leftarrow V \cup \{(e^*, w, \delta) : w \in \mathcal{D}, \delta \in \Delta\}$.

The above algorithm requires two tuning parameters: $M$, the maximum size of the model, and $o_{\max}$, the maximum order of interactions. Limiting the maximum order of interactions using $o_{\max}$ is an effective way of restricting the growth of $V$. Since with each addition of an element to the model, $C = |\mathcal{D}| \times |\Delta|$ elements are added to the variable space, with $o_{\max} = \infty$, the size of the variable space would be $|V| = C\tilde{\mathcal{E}} + |\mathcal{D}|$ at each updating step. Even though the variable space increases linearly with the model size, computational cost is still considerable for large $C$. Furthermore, multiple testing problems can become quite severe even with linear growth of $|V|$. We found that the algorithm works well when $\Delta$ is set to a small set of integers using prior knowledge about the ranges of different types of biological interactions, and $o_{\max}$ is set to 2 or 3. The setting of $\Delta$ is organism-dependent. We chose the values $\Delta = \{30, 100, 400, 1000\}$ for yeast and $\Delta = \{50, 200, 1000\}$ for Arabidopsis.

3.4. *Step* C. *Model pruning.* The stepwise procedure described in Step B results in a list of selected variables $\tilde{\mathcal{E}}$ of size $M$. The goal of the pruning step is to eliminate some of the "false positives" in $\tilde{\mathcal{E}}$ through backward deletion of variables. Selecting the best overall model depends upon the choice of a lack-of-fit function $lof(\cdot)$, which we describe in more detail below. First, we give the algorithm for backward deletion:

1. Let $E_M = \tilde{\mathcal{E}}$.
2. For $m = M - 1, \ldots, 1$, do

    (a) For $e \in E_{m+1}$, obtain model $E_m(e)$ by removing $e$ from $E_{m+1}$. Compute $lof[E_m(e)]$.
    (b) Let $e^* = \arg\min_j lof[E_m(e)]$.
    (c) Let $E_m = E_m(e^*)$, $lof_m = lof[E_m]$.

3. Pick $m^* = \arg\min_m lof_m$, and let $\mathcal{E} = E_{m^*}$.

We explored two different methods for assessing a model's lack of fit. The first is a weighted generalized cross validation error ($wGCV$). For a given model $\mathcal{E}$, let $d_{\text{reg}}(\mathcal{E})$ be the number of "regular" parameters in $\mathcal{E}$, which is equivalent to $|\mathcal{E}| + 1$. Let $d_{\text{knot}}(\mathcal{E})$ be the number of "knot" parameters, which is equal to the total number of distinct interactions in the model. Then, $wGCV$ is defined as the weighted sum of the GCV over each component:

$$wGCV(\mathcal{E}) = \sum_{j=1}^d \lambda_j^2 RSS_j / [1 - d(\mathcal{E})/G]^2,$$

where

$$d(\mathcal{E}) = d_{\text{reg}}(\mathcal{E}) + \gamma d_{\text{knot}}(\mathcal{E})$$

and

$$RSS_j(\mathcal{E}) = \|u_j - \hat{u}_j(\mathcal{E})\|^2$$

is the residual sum of squares of the least-squares fit of the model $\mathcal{E}$. In using $wGCV$, one needs to choose the smoothing parameter $\gamma$. Friedman (1991) discussed approaches for choosing this parameter for MARS, which is an adaptive model that also contains irregular knot parameters, and suggested using a fixed value of $\gamma = 2$ or a data adaptive value chosen through cross-validation. In our model, $\gamma$ represents the degrees of freedom of interactions of the form $e_{1,2} = (e_1, e_2, \delta)$. The choice of $\gamma$ needs to account for the maximization of the parameter $\delta$ over the set $\Delta$. For $e \in \mathcal{E}$, let $\tau_e = \sum_{g=1}^G I(\{X_g(e) > 0\})$ be the number of genes that contain at least one instance of $e$ in its promoter region. We use an adaptive rule of selecting a different $\gamma = \gamma(e_{1,2})$ for each interaction term $e_{1,2}$ through the following formula:

$$(7) \qquad \gamma(e_{1,2}) = 2[\log R + \log \tau_e + \log(G - \tau_e) - \log G]/\log G.$$

The intuition for the formula comes from the derivation of a modified Bayes information criterion for the model, described in Appendix A.2.

The second lack-of-fit criterion that we examine is a weighted version of the modified Bayes information criterion (wmBIC) given in Zhang and Siegmund (2007), which has the form $wmBIC(\mathcal{E}) = \sum_{j=1}^{d} \lambda_j^2 mBIC_j(\mathcal{E})$, where $mBIC_j$ for each component $j$ is defined as

$$
\begin{aligned}
mBIC_j(\mathcal{E}) \overset{\text{def}}{=} & \frac{1}{2}(G - d_{\text{reg}}(\mathcal{E}) + 1) \log \frac{RSS_j(\mathcal{E}_0)}{RSS_j(\mathcal{E})} \\
& + \log \frac{\Gamma[(G - d_{\text{reg}}(\mathcal{E}) + 1)/2]}{\Gamma[(G+1)/2]} \\
& + \frac{1}{2} d_{\text{reg}} \log RSS(\mathcal{E}_0) - \sum_{e \in \mathcal{E}} \log \tau_e + \log G - d_{\text{knot}} \log R.
\end{aligned}
$$

(8)

In the above formula, $\mathcal{E}_0$ is the model with only the intercept term. For derivation of this formula, see Appendix A.2 and Zhang (2005).

The aim of $wGCV$ is to reduce prediction error, with model parsimony as a secondary concern. In contrast, $wmBIC$ is derived under the Bayesian framework of maximizing posterior model probability instead of prediction accuracy. Hence, $wmBIC$, with a $\log n$ penalty for each degree of freedom, favors smaller models. In Section 4.1 we see that $wmBIC$ indeed selects a much smaller set of motifs than $wGCV$.

**4. Methods of validation.** As with all studies of this type, there is no single objective measure of performance. Experimental validation is the gold standard that is also hard to come by. In the absence of experimental validation, previous studies have relied on anecdotal evidence from existing literature, and some have used prediction error as a measure of performance. However, prediction error does not add to one's understanding of the model or interpretation of the results. For this reason, we employ three additional validation approaches, the second, based on flanking sequence analysis, is a new method.

A third method that we used to validate our results is gene list enrichment. If the motifs were true, then the set of genes that have that motif should be enriched with genes that are known to be related to the experiment. In Section 4.3 we describe the method we used for gene list enrichment analysis.

4.1. *Permutation analysis.* Permutation analysis allows us to compare results obtained using the real data with that performed on the randomly decoupled real data in which the genes' promoters are decoupled from their expression patterns and then re-associated at random.

The permutation procedure that we use is as follows. Let $\pi = (\pi_1, \ldots, \pi_n)$ be a random permutation of $(1, \ldots, G)$. Pair the expression vector $\mathbf{Y}_g$ of gene $g$ with the promoter sequence $\mathbf{S}_{\pi_g}$ of gene $\pi_g$. We call such a data set a randomly decoupled data set. The entire procedure detailed in Figure 1 (i.e., Steps A–C) is performed on this randomly decoupled data set. The *lof* curves from $N$ randomly decoupled data sets is compared to the *lof* curve of the real data set.

In Section 5.2 we show the results of applying this procedure to the Arabidopsis powdery mildew experiment. We obtained better results in yeast, which is a simpler organism with a higher signal to noise ratio.

4.2. *Flanking sequences.* An independent source of validation comes from the flanking sequences. If a promoter element found by our method were noise instead of signal, then the flanking sequences should not be any different from the background sequence. However, if the motif were in fact a real TFBS, then the flanking sequences may have a distribution with lower entropy that is different from the background sequence. This is because the promoter elements are composed of short words with which we hope to capture only the core consensus sequence, and binding sites often extend beyond the core sequence. This is especially true for transcription factors that bind over-represented sequences in the promoters of a particular genome, have binding sites with highly degenerate core sequences, or are members of large transcription factor families that bind a common core sequence (as is the case for many Arabidopsis transcription factors). Therefore, if the flanking sequences of a motif have lower entropy than their background sequences, then this is independent evidence that the motif is biologically significant.

Let the word $w$ be a component of the promoter element $e$. We find all locations of $w$ in $\mathbf{S}$ that appear as a component of some instance of $e$. Let $\mathcal{L}_{e,w}$ be the set of length $2L$ flanking sequences ($L$ bases on each side) of these appearances of $w$. Then, align the sequences in $\mathcal{L}_{e,w}$ to form the matrix $M_{e,w}$, where $M_{e,w}(i,a) = \sum_{l \in \mathcal{L}_{e,w}} I(\{l_i = a\})$. Also compute the background base frequencies $\{\pi_e(a) : a \in \mathcal{A}\}$ for promoters of genes in the set $\{g : X_g(e) > 0\}$. The sequence information content $I_{\text{seq}}$ of $M_{e,w}$ is defined as

$$I_{\text{seq}} = \sum_{j=1}^{2L} \sum_{i=1}^{A} M_{e,w}(i,j) \log \frac{M_{e,w}(i,j)}{\pi_e(i)}.$$

The statistical significance of a given value of $I_{\text{seq}}$ depends on the length of the flanking sequence $L$ and $N(p,w)$, which is the number of instances of $w$ that appear in the context of $e$. We will use the large deviations method developed by Hertz and Stormo (1999) for computing the p-value of $N(p,w)I_{\text{seq}}$.

4.3. *Gene list enrichment.* Gene list enrichment is our final method of validation. If the motifs identified were biologically real, then the set of genes that have that motif should be enriched in the half of the dataset exhibiting a differential pattern of expression compared with the control half of the dataset that did not exhibit a process-associated pattern.

Gene Ontology (GO) analysis is a popular method of statistical validation. If the set of genes that contain a motif is enriched for genes that belong to a GO-category that is related to the experiment, then this is evidence that the motifs are biologically meaningful. However, experiments usually perturb many pathways that relate to each other in a complex way. The genes in these pathways often belong to different GO categories, but may share common regulatory mechanisms. In addition, for some organisms, process-specific GO annotation is limited. To get around both of these issues, we use gene list enrichment as a validation method. If one has a list of genes whose expression is significantly impacted in the experiment (as we do), one can validate a given identified motif based on the probability of it being enriched in the promoters of genes whose expression was significantly impacted in the experiment compared with the entire data set. Let $N$ be the total number of genes used in the study, of which M belong to this list that contains all genes whose expression changes significantly in the experiment. Let $\tau_e$ be the number of genes that contain promoter element $e$, out of which $m_e$ appears in the list. The p-value for an observed value of $m_e$, based on the Fisher's exact test [Fisher (1922)], can be computed. A small p-value for this test is evidence that the element $e$ is a true regulatory element.

## 5. Results.

5.1. *Yeast.* We used the list of 1600 selected genes described in Section (2.1) for our analyses. The principal components of the $\alpha$-arrest experiments performed on these 1600 genes are given in Supplementary Figure 1 (the first 3 principal components are shown in Figure 2), with the scree plot given in Supplementary Figure 2 [Zhang, Wildermuth and Speed (2008)]. By the scree plot, there seems to be a drop in percent of variance explained from the third to fourth principal component. The first two principal components capture the periodic nature of the data, peaking respectively during the G2/M and M/G1 phases of the cell cycle. We choose as our basis set the first, second and third principal components, which together explain 63% of the total variance.

Table 1 gives a partial list of the promoter elements found for this data set. The complete list can be found in Supplementary Table 1(a) [Zhang, Wildermuth and Speed (2008)]. Of the 39 promoter elements in the model, 35 remain after backward deletion with BIC as the lack-of-fit criterion. Before deletion, the set of 39 motifs contain 7 singletons, 20 pairs and 12 triples.

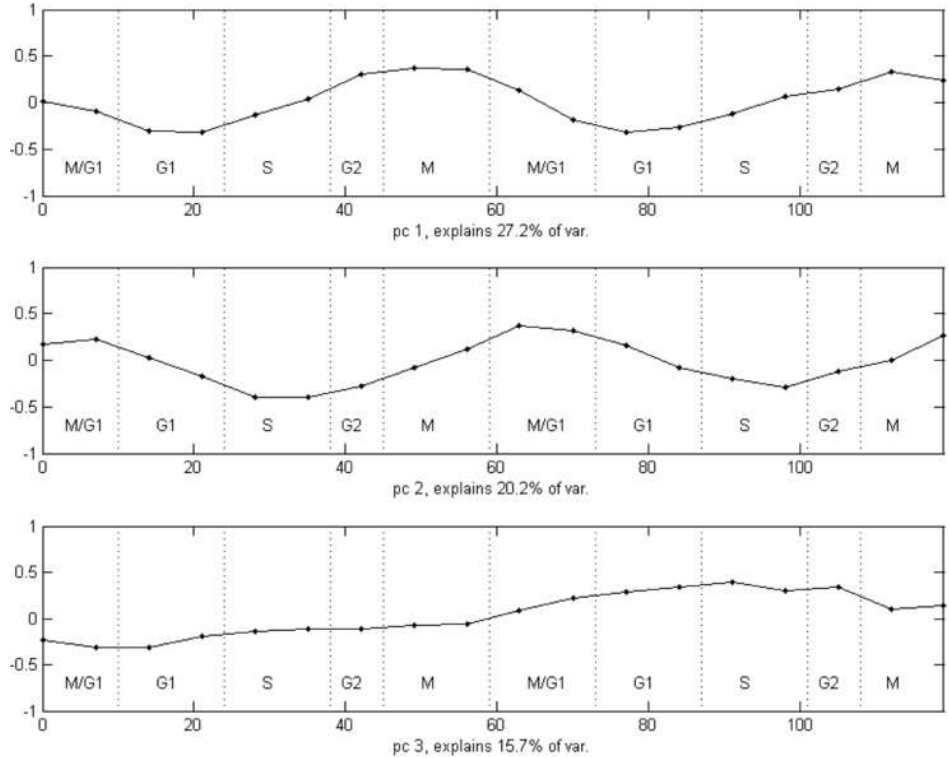

FIG. 2. *First three principal components of the Spellman et al. (1998) yeast cell cycle experiment.*

The table also shows, for each promoter element reported for the $\alpha$-arrest experiment, the number of genes that have that element that also belong to Spellman's 800 list. Note that Spellman's 800 genes comprise exactly 50% of the gene set on which we conducted our analysis. The gene list enrichment test results indicate that our method can extract promoter elements that are enriched in genes associated with the cell cycle.

Supplementary Table 1(b) [Zhang, Wildermuth and Speed (2008)] shows the flanking sequence analysis results for the yeast $\alpha$-arrest experiment. Note that most of the p-values are quite small (Figure 3 shows a histogram). This is strong evidence that many of the reported promoter elements are true positives.

Table 2 shows, for a small subset of the reported elements, the plots of the correlation coefficients between $X(e)$ and $Y_{t,\cdot}$ at each time point $t$, which we call "effect curves." We see that, because we considered projections on to principal components 1–3 simultaneously, we have found promoter elements that are influential for each of the different phases of the cell cycle. Of the 7 singleton motifs reported in Table 1, 5 are known motifs related to the

*Examples of promoter elements identified for the yeast α-arrest experiment. All of the elements mentioned in the case studies of Section 5.1 are shown here. The complete list can be found in Supplementary Table 1(a) [Zhang, Wildermuth and Speed (2008)]. Rank is reverse order of pruning from model, with the higher ranks being pruned first. "Putative site" is the assignment of known binding site names to the elements. If the element does not match exactly to any known motif, it is labeled newx, where x is the order of appearance in the list. Many of the "new" motifs are similar to and may be variations of known motifs. "Phase" is the phase of the cell cycle at which the effect of the promoter element is strongest. The columns n and m are the number of genes in the training set and in Spellman's 800 list, respectively, that contain the element. The column "p-value" contains the p-values for $(n, m)$ computed using Fisher's exact test*

| Rank | Motif | Putative site | Phase | $n$ | $m$ | p-value |
|------|-------|---------------|-------|-----|-----|---------|
| 1 | CGCGT | d-MCB | G1 | 582 | 380 | $8 \times 10^{-21}$ |
| 2 | ACGCGT | MCB | G1 | 180 | 141 | $1 \times 10^{-16}$ |
| 4 | TTTCGCG | SCB | mixed | 160 | 123 | $2 \times 10^{-13}$ |
| 8 | GCTGG | SWI5 | mixed | 809 | 400 | $7 \times 10^{-1}$ |
| 9 | TTGTTT | SFF | S/G2 | 1131 | 610 | $4 \times 10^{-7}$ |
| 12 | GGCTCCG | new8 | G2/M/G1 | 38 | 25 | $3 \times 10^{-2}$ |
| 14 | GCCCGTT | MCM1 | M | 59 | 27 | $8 \times 10^{-1}$ |
| 17 | (TGCTGGC,CGCGT,30) | SWI5, d-MCB | M/G1 | 7 | 7 | $8 \times 10^{-3}$ |
| 18 | (CGCGT,CGCGT,30) | d-MCB,d-MCB | G1 | 82 | 77 | $1 \times 10^{-18}$ |
| 21 | (TCGCGGG,TTGTTT,30) | new13, SFF | S, S/G2 | 5 | 5 | $3 \times 10^{-2}$ |
| 29 | (TTCGTGT,TTTCGCG,100) | SCB, SCB | G1 | 12 | 12 | $2 \times 10^{-4}$ |
| 31 | (TGGTCTG,TTTCGCG,400) | new19, SCB | S | 9 | 6 | $3 \times 10^{-1}$ |
| 35 | (TTTCCTA,TTGTTT,400) | MCM, SFF | M | 178 | 105 | $7 \times 10^{-3}$ |
| 37 | (TCCGAGC,CGCGT,100) | CSRE or GAL4, d-MCB | S | 9 | 7 | $9 \times 10^{-2}$ |
| 38 | (TGTTCTC,CGCGT,30) | new2, d-MCB | S | 7 | 7 | $8 \times 10^{-3}$ |

cell cycle (d-MCB, MCB, SCB, SWI5, SFF). The two "new" motifs are GGCTCCG and GCCCGTT. The latter, GCCCGTT, is a putative MCM1 site, because it aligns with the M phase and contains CCCGTT, which has been experimentally verified to be a MCM1 site in CLN3, SWI4 and CLB2. The pair-wise interactions are also very interesting. Below we list some noteworthy cases.

**MCM1–SFF pair:** Consider the motif (TTTCCTA,TTGTTT,400), rank 35. This motif combination appears in a large set of genes (178 total), most of which are categorized as M-phase genes by Spellman et al. (1998). Out of these 178 genes, 105 appear in Spellman's 800 list, which has a p-value of 0.007. These 178 genes include well-known players in the cell cycle, such as CDC10, SWI5, MCM3, SWI4, STE2, MCM6, STE3, STE6, CLB4, CDC5 and BUB2. TTTCCTA is a sub-word of the MCM1 binding site, while TTGTTT is the core of the SFF motif. This is strong evidence that this promoter element is a cooperative binding site for MCM1 and SFF. Since the MCM1 binding site is highly degenerate, we sough further evidence that TTTCCTA is part of an MCM1 site by analyzing its flanking region. The flanking region has information content $I_{seq} = 0.44$ for $N = 193$ instances, which has a highly significant p-value of $5 \times 10^{-12}$.

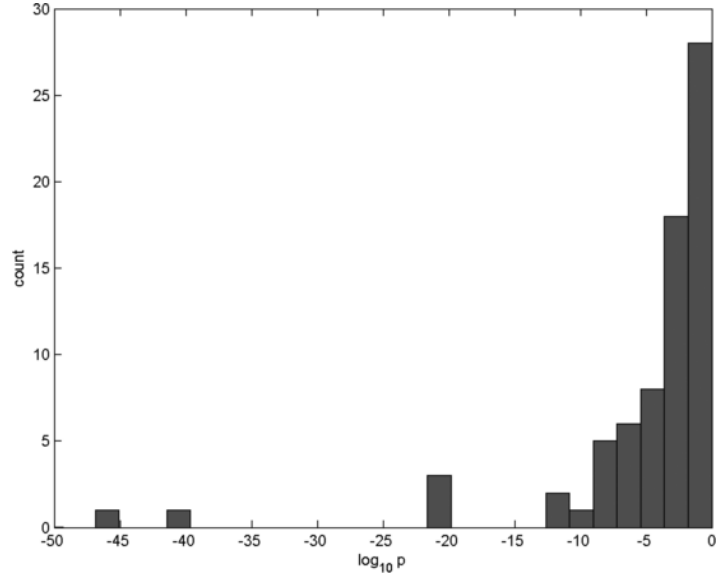

Fig. 3.  *Histogram for p-values of information content $NI_{seq}$ in flanking region of motifs discovered for yeast cell cycle. The flanking region of 20 bases (10 each on the left and right of the motif) is selected.*

**dMCB–dMCB pair:** (CGCGT,CGCGT,30), rank 18, is a short range interaction of two degenerate MCB motifs. Out of the 82 genes that have this element in their promoters, a highly significant 77 genes (p-value = $1 \times 10^{-18}$) are in Spellman's 800 list. This motif is strongly aligned with the G1 phase, which is consistent with existing knowledge about MCB activity. Comparing this with the element containing only MCB (row 1 of table), we see that by including the short-range interaction of CGCGT with itself, we can significantly reduce the number of false positive appearances. A GO analysis of the list of 82 genes that contains (CGCGT, CGCGT,30) returns significant hits to many GO categories, including DNA directed DNA polymerase activity. The list of 582 genes that contain CGCGT, however, is so diluted with many different functions that it is not significant for any one GO category. This shows that the distance-based interaction model captures additional relevant information. Finally, we look at the flanking sequences. $I_{seq}$ for CGCGT alone is already highly significant, with a value of 0.18 and a p-value of $6 \times 10^{-47}$. However, the $I_{seq}$ for (CGCGT,CGCGT,30) is far greater at 0.76, with a p-value that is essentially 0.

TABLE 2

*Plots of correlation coefficient between $X(e)$ and $Y_{t,g}$ for a selected set of promoter elements. The first column is the phase of the cell cycle during which the curve reaches its peak. The name and sequence of each binding site is in column 2. The vertical lines in the plot denote rough transition times between phases*

| Phase | Element | Effect curve |
|-------|---------|--------------|
| M/G1 | (TGCTGGC,CGCGT,30)<br>*SWI5, d-MCB* | 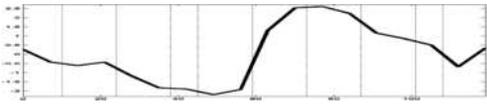 |
| M/G1 | GGCTCCG<br>*new8* | 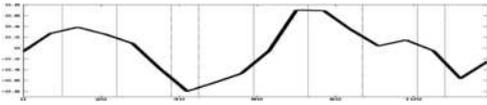 |
| G1 | (CGCGT,CGCGT,30)<br>*d-MCB, d-MCB* | 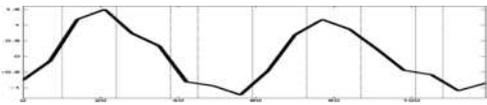 |
| G1 | (TTCGTGT,TTTCGCG,400)<br>*SCB, SCB* | 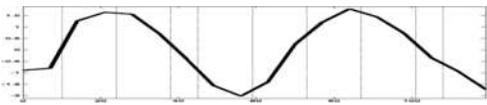 |
| S | (TCCGAGC, CGCGT,30)<br>*UAS1, d-SCB* | 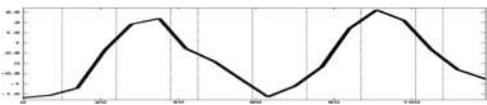 |
| S | (TGTTCTC, CGCGT,30)<br>*UAS1, d-MCB* | 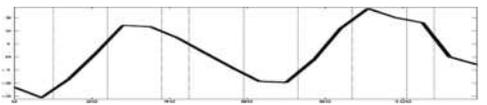 |
| S/G2 | TTGTTT<br>*SFF* | 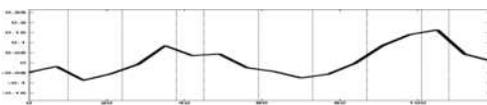 |
| M | (GCCCGTT)<br>*MCM1* | 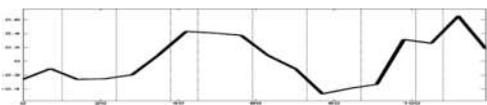 |
| M | (TTTCCTA,TTGTTT,400)<br>*MCM1, SFF* | 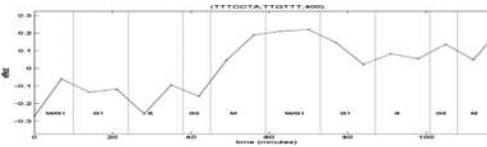 |

**SCB–SCB pair:** The element (TTCGTGT,TTTCGCG,100), rank 29, is an interaction of SCB motif TTTCGCG with the word **TTCGTG**T, which overlaps with T**TTCGTG**, also an instance of SCB. To verify that TTCGTGT is indeed part of an SCB motif, we analyzed its flanking sequence. The position that immediately precedes TTCGTG is indeed highly enriched for thymine. Thus, we hypothesize that (TTCGTGT,TTTCGCG,100) is a putative SCB–SCB pair. There are only 12 genes that contain this element, all of which belong to Spellman's 800 list, and all but 1 of them peak in G1. This list of 12 genes contain well-known cell cycle players CLN2, PCL1 and PCL2, and is enriched for the GO category cyclin dependent protein kinase regulator activity.

**SWI5–dMCB pair:** The first word in the element (TGCTGGC,CGCGT,30), rank 17, contains the SWI5 motif GCTGG, while the second word is dMCB. As seen from the plot of the effect curve, this motif is strongly aligned with M/G1 transition, which is consistent with the fact that SWI5 regulation occurs during this point of the cycle. A total of 7 genes contain this element, including the cell-cycle related transcription factors ASH1 and WTM1. This gene list is enriched for the GO categories hydrolase activity and beta-glucosidase activity.

**The histone clusters:** A striking result of applying our method to the yeast cell cycle experiment is that, without supervision, it was able to detect promoter elements associated with tightly regulated small sets of histone genes. These promoter elements (ranks 21, 31, 37, 38), along with the genes that have them, are listed in Table 3. Among the words contained in these promoter elements are the degenerate MCB motif CGCGT, the SFF motif TTGTTT, the SCB motif TTTCGCG, and some new motifs that are not commonly associated with the cell cycle: TGTTCTC, TCGCGGG, TGGTCTG and TCCGAGC. Among these "new" motifs, TTGTTCTC and TCCGAGC are parts of mapped UAS1/UAS2 elements [Osley (1991)].

5.1.1. *Comparison with previous methods.* All existing regression-based methods find motifs at each time point separately, and look across samples mainly for interpretation of already identified motifs. Therefore, in comparison to previous results, we emphasize that we do not expect to find exact concordance. The most significant motifs that we identified by cross sample analysis, such as SWI5, SFF, MCM1 and MCB, have also been identified by choosing the correct time point and using one of the previous single-sample methods [Bussemaker, Li and Siggia (2001), Das, Banerjee and Zhang (2004), Conlon et al. (2003) and Keles, Van der Laan and Vulpe (2004)]. However, as expected, there is no single time point that allows the identification of all of the motifs in our final set.

TABLE 3

*Promoter elements found for the yeast cell cycle experiment that are enriched with histone genes. For each promoter element, the genes that contain it are listed, along with their process and function annotations obtained from the Saccharomyces Genome Database (http://www.yeastgenome.org/)*

| Motif | ORF | YPD | Process | Function | Peak |
|-------|-----|-----|---------|----------|------|
| (TGTTCTC,CGCGT,30) | | | | | |
| | YBL002W | HTB2 | chromatin structure | histone H2B | S |
| | YBL003W | HTA2 | chromatin structure | histone H2A | S |
| | YDR224C | HTB1 | chromatin structure | histone H2B | S |
| | YDR225W | HTA1 | chromatin structure | histone H2A | S |
| | YGR014W | MSB2 | bud emergence | unknown | G1 |
| | YOR317W | FAA1 | fatty acid metabolism | long chain fatty acyl: CoA synthetase | M/G1 |
| | YPL127C | HHO1 | chromatin structure | histone H1 | S |
| (TCGCGGG,TTGTTT,30) | | | | | |
| | YBR009C | HHF1 | chromatin structure | histone H4 | S |
| | YBR010W | HHT1 | chromatin structure | histone H3 | S |
| | YDR261C | EXG2 | cell wall biogenesis | exo-beta-1,3-glucanase | S |
| | YKL096W | CWP1 | cell wall protein | beta-1,6-glucan acceptor | S/G2 |
| | YKL096W | CWP1 | cell wall protein | beta-1,6-glucan acceptor | S/G2 |
| (TGGTCTG,TTTCGCG,400) | | | | | |
| | YHR061C | GIC1 | bud emergence | binds Cdc42p | S |
| | YLR056W | ERG3 | sterol metabolism | C-5 sterol desaturase | S/G2 |
| | YNL030W | HHF2 | chromatin structure | histone H4 | S |
| | YNL031C | HHT2 | chromatin structure | histone H3 | S |
| | YOR247W | YOR247W | unknown | unknown; similar to Svs1p | G1 |
| | YPL111W | CAR1 | arginine metabolism | arginase | G2/M |
| | YLR162W | | | | |
| | YDR015C | | | | |
| | YNL323W | | | | |
| (TCCGAGC,CGCGT,100) | | | | | |
| | YBR009C | HHF1 | chromatin structure | histone H4 | S |
| | YBR010W | HHT1 | chromatin structure | histone H3 | S |
| | YDL037C | YDL037C | unknown | similar to glucan 1,4-alpha-glucosidase | G2/M |
| | YER001W | MNN1 | protein glycosylation | alpha-1, 3-mannosyltransferase | G1 |
| | YNL030W | HHF2 | chromatin structure | histone H4 | S |
| | YNL031C | HHT2 | chromatin structure | histone H3 | S |
| | YOR084W | YOR084W | unknown | unknown | G1 |
| | YER060W | | | | |
| | YER104W | | | | |

An interesting observation is that many of the previous methods found strong signals for motifs related to stress response [CCCCT and AGGGG in Bussemaker, Li and Siggia (2001)] and pheromone induction [STE12 motif in Conlon et al. (2003) and Das, Banerjee and Zhang (2004)]. These motifs are active in the first few time-points, and were hypothesized in Conlon et al. (2003) to be an experimental artifact due to centrifugation. We did not identify these motifs, because our approach uses linear projections to filter out processes that are not of interest.

Das, Banerjee and Zhang (2004) used MARS [Friedman (1991)] to find motif pairs in the yeast cell cycle data. Their model does not consider distance effects, but instead uses linear splines resembling a hockey stick to model what is hypothesized to be a switch-like behavior in gene transcription control. They used only the top 800 cell cycle related genes identified by Spellman et al. (1998), while we also included 800 control genes. They used simple degenerate words, as well as manually curated weight matrices. Their method also treats each time point separately. Thus, in finding subtle second order effects, one would expect significantly different models applied to different data to produce varying results illuminating different aspects of a complicated process. However, of the list of interacting motif pairs reported in Das, Banerjee and Zhang (2004), our results agreed by exact match in the motif pairs MCM1-SFF and SWI5-SFF, which are well-known pairwise interactions.

5.2. *Powdery mildew infection in Arabidopsis thaliana.* The Arabidopsis response to the pathogen involves multiple transcription factor family members. Though SA-dependent responses dominate systemic acquired resistance responses, ethylene (ET)- and jasmonic acid (JA)-dependent responses also play critical roles in the Arabidopsis response to the pathogen with outcomes dependent upon the complex interplay between these pathways. Table 4 shows the major transcription factors with known binding domain consensus sequences that are involved in the Arabidopsis defense response [e.g., review by Gurr and Rushton (2005)]. The three most well-studied of these transcription factors are the WRKY family- which mediate both SA- dependent and ET/JA-dependent responses [Ulker and Somssich (2004)], the ERF family, key regulators of ET- and JA-dependent defense-associated pathways [Gutterson and Reuber (2004)], and the TGA transcription factors which are able to interact with the SAR master regulator NPR1 [e.g., Johnson, Boden and Arias (2003)].

Figure 4 shows the first and second principal components of the powdery mildew experiment performed on the 3000 selected genes described in Section 2.2. The first principal component shows an increase in expression after infection in wild type plants with a reduced/delayed response in the *ics*1 mutant. It also included genes whose induced expression in wild type is abrogated in the *ics*1 mutant. These expression patterns reflect powdery mildew-induced genes that are partially or fully SA-dependent [Wildermuth et al. (2007)]. The second principal component is very interesting, as it shows an increase in expression over time in the ics1 mutant as compared to the wild type that exhibits little response to the pathogen. Genes that have a high score for this component may be involved in pathways that respond to the absence of ICS1 (and SA). Known genes associated with PC2 include genes associated with ET/JA-dependent responses [Wildermuth et al. (2007)]. We choose as our basis set the first and second principal components, which together explain 94.1% of the total variance.

TABLE 4
*Major transcription factors, and their corresponding binding sites, involved in Arabidopsis defense against the pathogen. Note that most of these transcription factors represent a multi-gene family*

| Factor name | Site name | Site consensus | Reference |
|---|---|---|---|
| WRKY | W-Box | (T)TGAC(T/C) | Eulgem (2005) |
| ERF | GCC-Box | (A)GCCGCC | Gurr and Rushton (2005) |
| TGA/OBF | as1/ocs | TGACG | Gurr and Rushton (2005) |
| MYC bHLH | G-Box | CACNTG | Gurr and Rushton (2005) |
| MYB | MYB | (T/C)AAC(T/G)G | Eulgem (2005) |
| | | G(G/T)T(A/T)G(G/T)T | Eulgem (2005) |
| SR genes | CGCG-Box | (A/C/G)CGCG(G/T/C) | Gurr and Rushton (2005) |

Table 5 gives a partial list of the motifs that were found by our method on this data set. The complete list can be found in Supplementary Table 2(a) [Zhang, Wildermuth and Speed (2008)]. The fourth column of the table shows whether the motif has a strong effect in the direction of principal component 1 or 2. A "strong effect" is declared if the list of genes that have that motif have a high ranking in the considered linear combination, with Wilcoxon rank-sum test p-value < 0.001. The table also lists the p-values of Fisher's exact test for enrichment in the top 1500 genes ranked by $\tilde{T}^2$ out of the total of 3000 genes in the filtered list (see Appendix A.1). Of the known TFBS, we found that most have strong effects along the first principal component, with the exception that the ERF binding site GCCGCC is aligned with the second principal component. Since the first principal component explains much more of the variance in the data set than the second principal component (80.7% compared to 13.4%), it is given a much larger weight in the model. Thus, most of the reported promoter elements aligned with the first principal component. The GCC-box, which has a very strong correlation with the second principal component, squeezed in to the list at number 42.

Our findings were validated in four ways. First, as shown in Table 5, we identified all six defense-associated motifs listed in Table 4 as well as the NFkB-like motif identified by Lebel et al. (1998). This motif is associated with innate immune responses in mammals, is present in the promoter of At-ICS1 [Wildermuth et al. (2001)], and has been implicated in plant response to pathogens. Second, we performed permutation analysis, the results of which are shown in Figure 5. Comparing the top and bottom plot, we see that *wmBIC* is indeed a more conservative model selection criterion than *wGCV*. The permutation results also show that our method decides upon a much larger model for the real data set than for the randomly decoupled data sets. This fact gives confidence that some of the discovered motifs for the real data set are biologically meaningful. Third, we analyzed the flanking regions of the identified motifs and presented the p-values associated with this analysis in Supplementary Table 2(b) [Zhang, Wildermuth and Speed (2008)]. Overall, these p-values were less significant than those for the yeast motifs, likely due to the presence of large transcription factor families that bind a similar core motif in Arabidopsis. However, a number of the identified motifs did exhibit significant conservation of flanking sequences. Finally, the results of our gene list enrichment analysis is shown in Table 5 . We include a detailed discussion for a few interesting case studies extracted from the results below.

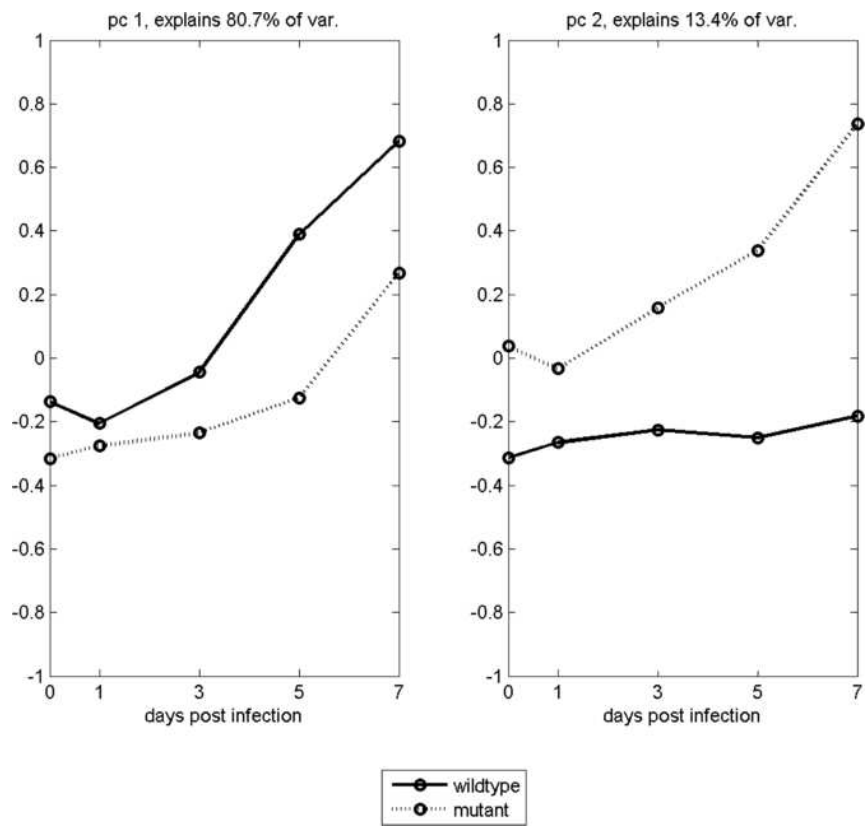

FIG. 4.   *First two principal components of the Arabidopsis powdery mildew infection experiment.*

**W-box clusters and interactions:** WRKY transcription factors are critical regulators of plant response to abiotic and biotic stress [Ulker and Somssich (2004)]. WRKY transcription factors bind the W-box core TGAC, with reported flanking sequences in Arabidopsis involved in plant defense response biased toward TTGAC(T) [e.g., Yu, Chen and Chen (2001)] and TTGACC [Laloi et al. (2004)].

W-box containing elements are significantly aligned with both the first and second principal components. This makes sense biologically, as WRKY factors can modulate both SA-dependent and ET/JA-dependent path-

TABLE 5

*Examples of promoter elements identified for the Arabidopsis powdery mildew infection experiment, listed in reverse order of pruning from the model. All of the elements mentioned in the case studies of Section 5.2 are shown here. The complete list can be found in Supplementary Table 2(a) [Zhang, Wildermuth and Speed (2008)]. "Putative site" is the assignment of known binding site names to the elements. Putative sites listed in Table 4 are specified by name as is the NFkB-like motif identified by Lebel et al. (1998) and associated with innate immunity. All other identified motifs are listed as new, though some of these exhibit significant overlap with known motifs [Higo et al. (1999)]. "Component" is the principal component with which the element has a strong effect, and "none" if no principal component significantly dominates the other. The columns $n$ and $m$ are the number of genes in the training set and in the top 1500-list, respectively, that contain the element. The column "p-value" contains the p-values for $(n, m)$ computed using Fisher's exact test*

| Rank | Motif | Putative site | Component | $n$ | $m$ | p-value |
|------|-------|---------------|-----------|-----|-----|---------|
| 1 | GACTTT | NF$\kappa$B-like | 1,2 | 1672 | 909 | $5 \times 10^{-8}$ |
| 2 | TTGACT | W-box | 1,2 | 1629 | 914 | $2 \times 10^{-13}$ |
| 4 | (TGACTA,TTGACC,1000) | W-box, W-box | 1 | 445 | 268 | $2 \times 10^{-6}$ |
| 5 | AGTCTT | NF$\kappa$B-like | 1,2 | 1460 | 802 | $9 \times 10^{-8}$ |
| 8 | TGACGT | TGA | none | 711 | 398 | $2 \times 10^{-4}$ |
| 11 | (AGACTT,TTGACT,200) | NF$\kappa$B-like, W-box | 1,2 | 478 | 313 | $8 \times 10^{-14}$ |
| 12 | (GTCGTC,TTGACT,200) | new6, W-box | 1 | 197 | 130 | $2 \times 10^{-6}$ |
| 19 | (CATGTG,GAAATA,1000) | Myc, new11 | 1 | 665 | 320 | $9 \times 10^{-1}$ |
| 21 | (TTCGTC,TTGACT,200) | new15, W-box | 1 | 293 | 192 | $1 \times 10^{-8}$ |
| 25 | (CGCGTT,TTTCCA,200) | CGCG-box, new8 | 1 | 72 | 42 | $9 \times 10^{-2}$ |
| 26 | (TCAAAC,TTGACC,200) | new19, W-box | 1 | 366 | 210 | $2 \times 10^{-3}$ |
| 28 | (GAGCTT,TTGACC,1000) | new20, W-box | none | 376 | 200 | $1 \times 10^{-1}$ |
| 29 | TCAACG | Myb | 1 | 951 | 556 | $2 \times 10^{-10}$ |
| 33 | (TGTCGA,TTGACC,200) | new23, W-box | none | 106 | 59 | $1 \times 10^{-1}$ |
| 42 | (GGCGGC,AATTTT,200) | GCC-box, new3 | 2 | 145 | 76 | $3 \times 10^{-1}$ |

ways, the latter of which are likely to be associated with principal component 2.

In our analyses, the TTGACT motif appeared alone (Rank 2), and in close proximity (within 200 nt) to other motifs: NFκB-like motif AGACTTT (rank 11), TCAACT (rank 12) and TTCGTC (rank 21). In addition, it appears within 1000 nt of TGACTA, which contains the W-box core (Rank 4). Enrichment of W-boxes in promoters of genes with altered expression in response to biotic stress is consistently observed [e.g., Maleck et al. (2000)], in agreement with our finding of a W-box, W-box pair.

The TTGACC appeared in combination with other motifs in close proximity [TCAAAC (rank 26) and TGTCGA (rank 33)] or within 1000 nt [TGACTA (rank 4) and GAGCTT (rank 28)]. In all of these instances, the flanking sequences of these W-boxes have p-values $< 0.001$. Resolving flanking sequence specificity and genes targeted by specific WRKY factors has been extremely challenging as the Arabidopsis WRKY family has over 70 members and promoters of regulated genes tend to contain multiple W-boxes. Furthermore, stimulus-dependent changes in WRKY binding affinities result in WRKY shuffling on promoter elements [Turck, Zhou and Somssich (2004)]. Our approach allows us to identify putative WRKY factor interaction pairs *in silico* and to predict those cases where flanking sequences may be more readily resolved, greatly facilitating experimental efforts.

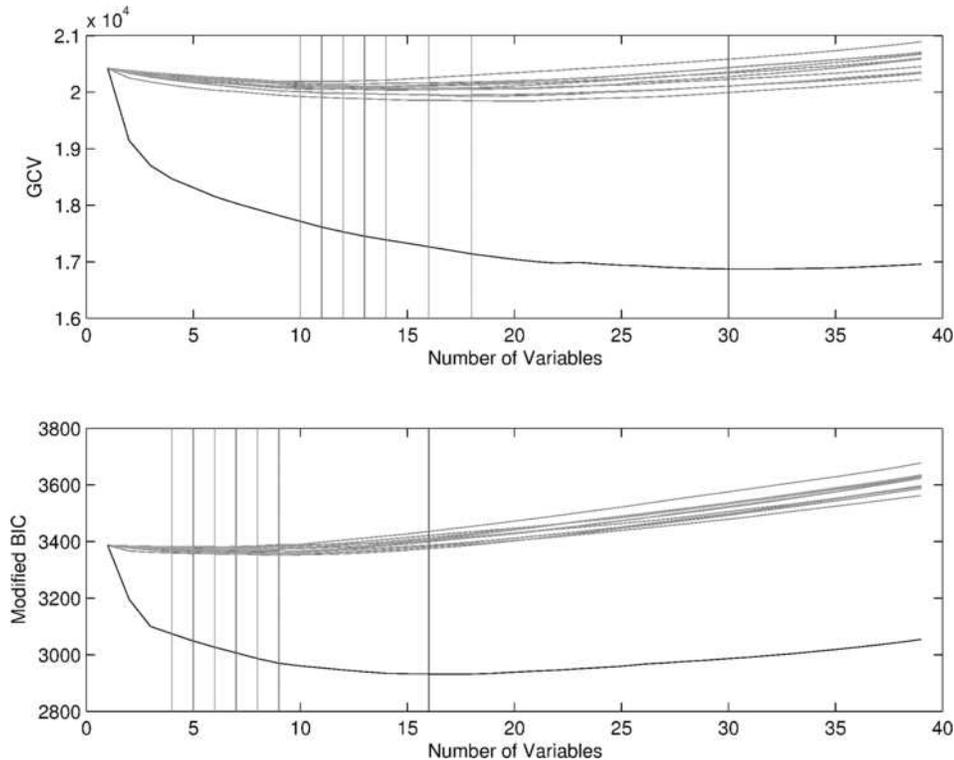

Fig. 5. *Permutation analysis for the Arabidopsis powdery mildew infection data. The top and bottom plots show, respectively, the wGCV and wmBIC curves for 10 randomly decoupled data sets (gray lines) versus the real data set (black line). Vertical lines show the model size that minimizes these curves.*

**(CGCGTT, TTTCCA, 200)** This motif combination (ranked 25) appears in a set of 72 genes. This set of 72 aligns significantly with principal component 1. When we look at the flanking sequences, $I_{seq}$ for CGCGTT is highly significant with a value of 0.76 and a p-value of $5 \times 10^{-6}$. Interestingly, the TTTCCA motif also appears in another motif combination (TCAACT, TTTCCA, 200) ranked 13. In this case, analysis of the flanking sequence for TTTCCA is 0.14, with a p-value of $2 \times 10^{-6}$.

The CGCGTT motif comprises part of a known motif, the "CGCG box," with consensus sequence (A/C/G)CGCG(C/G/T); this CGCG box is recognized by all 6 members of the *A. thaliana* signal responsive genes AtSR1-6 [Yang and Poovaiah (2003)]. The Arabidopsis SR proteins are Ca2+/calmodulin-binding/DNA-binding proteins that are induced in response to a variety of plant phytohormones and stresses [Yang and Poovaiah (2003)]. Ca2+ plays an important role in mediating SA and H2O2 signal transduction [Yang and Poovaiah (2003)]; however, our knowledge about the specific mechanisms involved is limited. AtSR3-6 have been found to be rapidly induced in response to treatment by salicylic acid or H2O2 [Yang and Poovaiah (2003)]. We do not observe statistically significant changes in expression for any of the AtSR genes over the time course of powdery mildew infection. However, as this time course focused on later stages of infection (1–7 days), it is very possible that early transcriptional responses (such as a possible change in AtSR transcription) were not captured. Instead, we resolve the later progressive transcriptional response associated with extensive growth and reproduction of the powdery mildew including downstream genes (e.g., containing the CGCG box) that may be regulated by rapidly-induced transcription factors such as the AtSRs. Though the 72 genes with the (CGCGTT, TTTCCAA, 200) motif combination were not significantly enriched in any MIPS GO functional annotation category (performed using Virtual Plant 0.9, BioMaps function), this 72 gene set includes a number of transcription factors, defense-related genes and a calcium transporting ATPase (see Supplementary Table 3) [Zhang, Wildermuth and Speed (2008)]. To further assess whether additional members of the 72 gene set had been previously found to be directly modulated by Ca2+/calmodulin, we compared the 72 gene set with a union of genes (of 709) compiled from the following Arabidopsis datasets: (1) the AtSR genes (6 genes) and genes identified as containing the CGCG box (19 genes) [Yang and Poovaiah (2002)]; (2) the Ca2+/calmodulin-binding, BTB and TAX domain-containing AtBT protein family (5 genes) and interactors (2 genes) [Du and Poovaiah (2004)]; (3) rapid calcium-responsive up and down regulated genes (229 genes) [Kaplan et al. (2006)]; (4) calmodulin-binding proteins identified using high density protein arrays (173 genes) [Popescu et al. (2007)]; and (5) genes whose annotation included the keyword calcium or calmodulin (303 genes), obtained using VirtualPlant 0.9. Of the 72 genes in our combined motif set, only 3 were present in the compiled Ca2+/calmodulin gene set (see Table 1). This suggests that we may have elucidated a previously uncharacterized specific subset of responses requiring a CGCGTT *cis*-acting element and TTTCCAA element in close proximity that can then be experimentally validated.

**AATTTT, GGCGGC** To our knowledge, the AATTTT motif has not previously been described in its entirety as a *cis*-acting regulatory element in Arabidopsis. The AATTTT motif is a component of the plant *cis*-acting regulatory element CAAAATTTTGTA [PLACE database motif S000466, Higo et al. (1999)] and is specifically activated during the early phases of an incompatible plant/bacterial pathogen interaction in tobacco [Pontier et al. (2001)].

It appears alone (rank 7) and in combination with other motifs. For ranks 18, 36 and 42, the AATTTT motif is within 200 nt of its partner; whereas for ranks 22 and 41, it is within 1000 nt of the other motif. The p-values for the $I_{seq}$ values AATTTT in these interactions are all quite low. The (GGCGCC, AATTTT, 200) pair is especially interesting as it is the only element that is mainly associated with the second component.

Genes associated with component 2 exhibit a trend of enhanced expression in the SA biosynthetic mutant compared with wild type over the time course of expression and may be negatively regulated by SA. The GGCGCC motif is commonly known as the GCC box recognized by ethylene-responsive factors (ERFs). ERF transcription factors regulate developmental and defense processes and are associated with ET and JA signal transduction pathways [Gutterson and Reuber (2004)]. The set of genes with the GCC motif is 245; this set is highly ranked when component 2 alone is examined, but falls below our threshold when the loss function combines both components 1 and 2. The GCCGCC set contains the defensin PDF1.2 (At5g44420), a marker of ET- and JA acid-dependent defense responses, regulated in part by ERFs. The (GGCGCC, AATTTT, 200) pair does not include PDF1.2, but does include the ERF At1g06160 and AtWRKY75 (At5g13080) transcription factors, as well as defense-related genes associated with ET- and JA-dependent defense responses such as chitinases, a germin-like protein, and defensin-fusion protein (At2g26020). Both ERF and WRKY factors can mediate cross talk between SA- and ET/JA-dependent signaling pathways. This is particularly interesting as this subset includes component 1- (SA-dependent responses) and component 2-associated genes.

**6. Discussion.** We have shown using two experimental data sets that the model and methods we propose in Section 3 can be quite useful in finding transcription factor binding sites using multivariate gene expression data. The model stated in (1) and (3) can be quite general to accommodate any linear contrast(s) of interest. For cases where no obvious contrasts are available from the experimental design, we suggest selecting the basis vectors $\{\mathbf{v}_j\}$ using principal components. For both the yeast $\alpha$-arrest experiment and the Arabidopsis powdery mildew infection experiment, the first few principal components are very effective in capturing meaningful structure in the data.

To model cooperative regulation between TFBS, we developed a recursive model for interactive effects that is limited to a chosen range along the promoter sequence. The range parameter is also chosen during the model fitting process, and a model selection criteria is proposed to adjust for this additional degree of freedom. However, these model selection criterion are only meant as a guideline for interpreting the models, and should not be taken as strict rules for inclusion and exclusion of variables.

In addition to validation using published experimental literature, we employed three simple methods for interpretation of the model and statistical validation of the reported motifs. We found the permutation procedure to be quite useful (and necessary) for assessing how much noise the approach is expected to approach. The flanking sequence and gene list enrichment analyses are particularly important to experimental biologists as it allows them to prioritize motifs of interest and to understand the differential variability in flanking sequences of particular motifs.

## APPENDIX

**A.1. Data pre-processing.**

A.1.1. *Yeast $\alpha$-arrest experiment.* As in Spellman et al. (1998) and Bussemaker, Li and Siggia (2001), we limit our search of TFBS to the 700 base promoter sequence upstream of the gene immediately preceding the transcription start site. Missing data have been imputed using KNNimpute [Troyanskaya et al. (2007)]. Prior to the analyses, the gene expression values from each sample were centered and scaled to have mean 0 and variance 1.

We chose the genes for our analysis as follows: The list of 800 "cell-cycle related" genes identified by Spellman et al. (1998) are automatically included (we refer to these as Spellman's 800 in our analysis). To choose the negative controls, we first clustered the data using K-means, and then identifying those clusters that by visual examination did not exhibit a cell-cycle related pattern (these genes have very little variation across the 18 time-points). 800 genes are sampled randomly from these clusters to be included in the reduced set. Thus, we used 1600 genes in our final analysis, exactly 50% of which are in Spellman's 800 list.

A.1.2. *Powdery mildew infection experiment.* *Arabidopsis thaliana* Columbia strain wildtype and *ICS*1-null mutant plants were evenly positioned and intermixed in flats consisting of 6 boxes and placed in growth chambers [Wildermuth et al. (2007)]. Each box contained 12 plants. When the plants were four weeks old, they were infected with a heavy innoculum of powdery mildew (*Golovinomyces orontii*). Uninfected plants served as controls and were grown in growth chambers with identical conditions. Mature leaves were harvested (for RNA) at 6 time points: (0 hr, just prior to infection), 6, 24, 72, 120 and 168 hpi. Plants could not be re-sampled, so at each time point, paired samples were harvested from two randomly selected plants. mRNA extraction, target labeling and hybridization to Affymetrix Arabidopsis ATH1 GeneChips was performed for 4 complete biological replicates, yielding information on 22810 probesets for 56 arrays. For our analysis, we averaged the expression level in the 4 biological replicates for each gene, time point and plant type (mutant or wild type).

We discarded data points at hour 6, due to the fact that they were not collected at the same time during the day as the other samples and thus, circadian effects, rather than effects due to infection, could confound our analysis. Out of the 22810 probesets, we selected smaller, filtered gene sets for further analysis using the $\tilde{T}^2$-statistic from Tai and Speed (2006). $\tilde{T}^2$ is an empirical Bayes statistic that ranks genes from replicated time-course experiments by differential expression over time in a single biological condition or across multiple biological conditions. Our filtered gene set contains the top 1500 and bottom 1500 genes ranked by the $\tilde{T}^2$ statistic for differential expression over time between the wildtype and *ics*1-mutant strains. The bottom 1500 genes included in each gene list are necessary as negative controls.

For $\mathbf{S}_g$, we extracted the 1000 base promoter sequence upstream of gene $g$ immediately preceding the translation start site. Thus, $\mathbf{S}_g$ includes the 5′ UTR sequence.

**A.2. Model selection criterion.** For ease of notation, we first assume that there is only one basis vector, and thus, the responses $\mathbf{Y}_g$ are univariate for each gene $g$. Also, for simplicity, we assume that the variance of the error term $\epsilon_g$ in (3) is known and equal to 1 [the unknown variance case yields similar degrees of freedom calculations, see Zhang (2005)]. Let $e_1$ and $e_2$ be two promoter elements. By the notation of Section 3.3, $A_g(e_1)$ and $A_g(e_2)$ are the locations of $e_1$ and $e_2$, respectively, in $\mathbf{S}_g$. Define

$$D_g(e_1, e_2) = \min\{|k - l| : k \in A_g(e_1), l \in A_g(e_2)\}$$

to be the minimum distance between any pair of $(e_1, e_2)$ in $\mathbf{S}_g$. For $A_g(e_1)$, $A_g(e_2)$ empty, $D_g(e_1, e_2)$ is defined to be $\infty$. Let $e_{1,2} = (e_1, e_2, \delta)$ be the promoter element representing the $\delta$-range interaction of $e_1$ and $e_2$. Then, by the definition of $X(e)$ in (6), inclusion of $e_{1,2}$ adds the term $\alpha I(D_g(e_1, e_2) < \delta)$ to the existing model. That is, the model that includes $X(e)$ can be re-written as

$$(9) \qquad Y_g = \mu + \sum_{e \in \mathcal{E} \setminus e_{1,2}} \beta(e) X_g(e) + \alpha I(D_g(e_1, e_2) < \delta) + \epsilon_g,$$

where $\delta$ is a change-point parameter. Including $e_{1,2}$ as a predictor adds the parameters $\gamma$, $\delta$ to the model. We give here a crude analysis of the effective degrees of freedom contributed by this interaction term.

We assume that the promoter length $R$ and the prior for $\delta$ are fixed and do not increase with the sample size $G$, so that the probability

$$\pi_\delta \equiv P(D_g(e_1, e_2) < \delta)$$

can be considered as a fixed function of $\delta$. Without loss of generality assume that the gene indices are ordered so that $D_g(e_1, e_2)$ is monotone nondecreasing. Define

$$\tau_\delta = \max_i\{D_i(e_1, e_2) \leq \delta\}$$

to be the number of genes that have the element $e_{1,2}$. Then, assuming that the promoter sequences $\mathbf{S}_g$ are i.i.d., $\tau_\delta$ is binomial with log-likelihood

$$\log P_\delta(\tau_\delta = m) = \log \binom{G}{m} \pi_\delta^m (1 - \pi_\delta)^{G-m}$$

$$= \frac{1}{2} \log \frac{G}{2\pi m(G - m)} - G I_\delta,$$

where $I_\delta$ is the large deviations constant

$$I_\delta = \frac{m}{G} \log \frac{G \pi_\delta}{m} + \frac{G - m}{G} \log \frac{G \pi_\delta}{G - m}.$$

For ease of computation, we assume that $e_{1,2}$ is the only term in the model, and thus, (9) is reduced to $Y_g = \mu + \alpha I(D_g(e_1, e_2) < \delta) + \epsilon_g$. Let $\mathcal{M}_0$ be the model where $\alpha = 0$, and $\mathcal{M}_1$ be the alternative model where $\alpha$ is arbitrary. Then, under the Bayesian model selection framework, we choose the model with the largest Bayes factor, which has the form

$$
\frac{P(\mathcal{M}_1|Y)}{P(\mathcal{M}_0|Y)}
$$

$$
= \left( \int_0^R \int_{-\infty}^{\infty} \int_{-\infty}^{\infty} \exp\left( -\frac{1}{2}\left[ \sum_{i=1}^{\tau_\delta} (Y_i - \alpha - \mu)^2 \right. \right. \right.
$$

$$
\left. \left. + \sum_{i=\tau_\delta+1}^{G} (Y_i - \mu)^2 \right] \right.
$$

$$
\left. \left. + \log P_\delta(\tau_\delta) \right) d\alpha \, d\mu \, d\delta / R \right)
$$

$$
\times \left( \int_{-\infty}^{\infty} e^{-(1/2)\sum_i (Y_i - \mu)^2} \, d\mu \right)^{-1}.
$$

(10)

In (10) we have assumed uniform priors for $\delta$, $\mu$ and $\alpha$.

Since $\pi_\delta$ does not change between $\mathcal{M}_0$ and $\mathcal{M}_1$, the term $GI_\delta$ in the numerator of (10) converges to a chi-square distribution under both models, and thus is stochastically bounded away from 0 and $\infty$ as $G \to \infty$. Also, due to this assumption, with probability one, $\tau_\delta = O(G)$, and thus, the methods in Zhang and Siegmund (2007) can be directly applied to the evaluation of (10) to yield the following approximation when $G$ is large:

$$
\log \frac{P(\mathcal{M}_1|Y)}{P(\mathcal{M}_0|Y)} = l(\hat{\alpha}, \hat{\mu}, \hat{\delta}) - [\log(\tau_\delta) + \log(G - \tau_\delta) - \log G]
$$

$$
- \log R + O_p(1),
$$

(11)

where $l(\hat{\alpha}, \hat{\mu}, \hat{\delta})$ is the maximized likelihood. Compared with the classic BIC which has a penalty of $\frac{1}{2} p \log G$, where $p$ is the degrees of freedom of the model, this new result suggests that each interaction term $(\delta, \gamma)$ contributes

$$
2[\log R + \log \tau_\delta + \log(G - \tau_\delta) - \log G]/\log G
$$

(12)

degrees of freedom to the model.

When the response is multivariate with weights $\{d_j\}$ as in (5), we simply take a weighted sum of the BICs for each component.

The variance unknown case is more technically messy, but the same logic applies, yielding approximation (8). The proof will not be shown here; the interested reader can refer to Zhang (2005).

## SUPPLEMENTARY MATERIAL

**Additional tables and figures** (doi: [10.1214/07-AOAS142SUPP](10.1214/07-AOAS142SUPP); .zip). Supplementary Figures 1 and 2 show respectively the principal components and screeplot for Spellman et al. (1998) yeast cell cycle data set. Supplementary Table 1 (a-b) shows gene list enrichment, annotation, and flanking sequence analysis for promoter elements identified in Spellman et al. (1998) yeast cell cycle experiment. Supplementary Table 2 (a-b) shows the same information for the Wildermuth et al. (2007) arabidopsis powerdery mildew infection experiment. Supplementary table 3 lists the genes containing the (CGCGTT, TTTCCA, 200) element.

N. R. ZHANG
DEPARTMENT OF STATISTICS
STANFORD UNIVERSITY
STANFORD, CALIFORNIA 94305-4065
USA
E-MAIL: nzhang@stanford.edu

M. C. WILDERMUTH
DEPARTMENT OF PLANT AND MICROBIAL BIOLOGY
UNIVERSITY OF CALIFORNIA
BERKELEY, CALIFORNIA 94720-3102
USA
E-MAIL: wildermuth@nature.berkeley.edu

T. P. SPEED
DEPARTMENT OF STATISTICS
UNIVERSITY OF CALIFORNIA
BERKELEY, CALIFORNIA 94720-3860
USA
E-MAIL: terry@stat.berkeley.edu